\documentclass[preprint]{aastex}
\shorttitle{DIFFUSIVE PROTOPLANET MIGRATION}
\shortauthors{JOHNSON, GOODMAN \& MENOU}

\newcommand\au{\mbox{\,\textsc{au}}}
\newcommand{\yr}{\,{\rm yr}}
\begin{document}
\title{Diffusive Migration of Low-Mass Proto-planets in Turbulent
Disks}

\author{Eric T. Johnson and Jeremy Goodman}
\affil{Department of Astrophysical Sciences,
       Princeton University, Princeton, NJ 08544}
\and
\author{Kristen Menou}
\affil{Department of Astronomy, Columbia University, 
550 W. 120th St., New York, NY 10027}

\begin{abstract}
  Torque fluctuations due to magnetorotational turbulence in
  proto-planetary disks may greatly influence the migration patterns
  and survival probabilities of nascent planets. Provided that the
  turbulence is a stationary stochastic process with finite amplitude
  and correlation time, the resulting diffusive migration can be
  described with a Fokker-Planck equation, which we reduce to an
  advection-diffusion equation. We calibrate the coefficients with
  existing turbulent-disk simulations and mean-migration estimates,
  and solve the equation both analytically and numerically. Diffusion
  tends to dominate over advection for planets of low-mass and those
  in the outer regions of proto-planetary disks, whether they are
  described by the Minimum Mass Solar Nebula (MMSN) or by T-Tauri
  alpha disks.  Diffusion systematically reduces the lifetime of most
  planets, yet it allows a declining fraction of them to survive for
  extended periods of time at large radii.  Mean planet
  lifetimes can even be formally infinite (e.g. in an infinite steady
  MMSN), though median lifetimes are always finite. Surviving
  planets may linger near specific radii where the combined effects of
  advection and diffusion are minimized, or at large radii, depending
  on model specifics. The stochastic nature of migration in turbulent
  disks challenges deterministic planet formation scenarios and
  suggests instead that a wide variety of planetary outcomes are
  possible from similar initial conditions.  This would contribute to
  the diversity of (extrasolar) planetary systems.
\end{abstract}


\section{Introduction}

A decade of extrasolar planet discoveries has shown that the process
of planet formation is more complex than originally anticipated. It
leads to a remarkable diversity of planetary configurations, ranging
from migrating hot Jupiters to eccentric giant planets, as well as our
own ``circular'' Solar System.  Following these discoveries, 
progress has been made in understanding planet formation, but the 
theory is still incomplete \citep[e.g.][]{mcm00,wgl00,pt06,ar05}.

The leading scenario for the formation of giant planets is the core
accretion mechanism. Icy planetesimals located beyond the snow line in
their host disk collide repeatedly to grow a core with a modest
gaseous atmosphere. If this core succeeds in reaching a critical mass
of a few tens of Earth masses, runaway accretion of a massive gaseous
envelope proceeds and leads, ultimately, to the formation of a gaseous
giant planet \citep{saf69,miz80,pol96}. Some evidence supporting this
scenario has emerged in recent years, in the form of a metallicity
trend for stars hosting planets \citep[e.g.][]{san04,gil00,wel05}, the
discovery of a high density hot Jupiter \citep{sat05} and that of a
surprisingly low-mass micro-lensing planet \citep{beau06}.  However, a
long-standing difficulty for the core accretion scenario, which has
not yet been fully elucidated, is the fact that the timescale required
to build-up critical core masses and thus large gaseous envelopes
($\sim 10^6$-$10^7$~yr) is comparable to the lifetimes of
proto-planetary disks \citep[e.g.][]{pol96,pt99}.

Orbital migration adds a layer of complication to theories of planet
formation.  As a result of gravitational interactions with their
gaseous disk \citep{gt80,lp86}, the orbits of planets in the
terrestrial mass range are predicted to decay on timescales ($\sim
10^5$~yr) short compared to disk lifetimes
\citep[e.g.][]{kp93,w97a,w97b}.  Migration is slower for planets of
much smaller or much larger masses: in the first case because the
torque causing migration is quadratic in planet mass, and in the
second case because the planet opens a gap and then migrates on the
disk's accretion timescale, which can be comparable to its lifetime.
While it is possible or probable that many terrestrial planets form by
agglomeration of smaller bodies after the gas is gone, this is not an
option for the solid cores of Jovian planets since, in the core
accretion scenario, the cores must form before the gaseous envelopes
can be accreted.  The prevalence of Jovian planets with orbital
periods of only a few days deepens the mystery as it suggests that
these planets did migrate but stopped short of merging with their
stars at orbital radii where even the accretion timescale would seem
to have been very short \citep[e.g.][]{lbr96}.

Analytic calculations and most hydro-dynamical simulations of
migration usually assume a disk that is laminar apart from the waves
and shocks excited by the planet itself
\citep[e.g.][]{kp93,w97a,kdh01,nb03a,nb03b,dkh03,luf03,sch04}.  But
the effective viscosity of disks probably involves turbulence.
\citet{np04}, \citet{lsa04}, and \citet[hereafter N2005]{nelson05}
have found in 3D simulations of magneto-rotational turbulence that the
instantaneous torque exerted on a planet in the terrestrial mass range
is subject to fluctuations many times its mean value, apparently
caused by turbulent density fluctuations in the planet's vicinity. In
fact, no obvious secular decay manifests itself in the orbits of
planets with mass $M_p\lesssim 10 M_\oplus$, although because the
simulations are limited to $\sim 10^2$ planetary orbits---compare
$\gtrsim 10^5$ for Jupiter during the lifetime of the proto-solar
nebula---and the predicted decay in semimajor axis is $\lesssim 10\%$
over this period, the expected secular trend might be difficult to
discern amid the fluctuations.

The possibility that planetary torques are stochastic raises several
issues:

\begin{enumerate}
\item Is the time-averaged torque exerted on a planet of a given
angular momentum affected?  N2005 finds that turbulence excites
planetary eccentricity, $e$. In linear theory, eccentricity allows the
planet to interact with the disk at additional resonances that do not
couple to circular orbits \citep{gt80,gs03}.  Also, at second and
higher orders in $e$, the radial positions of the original resonances
shift if one compares eccentric and circular orbits at fixed angular
momentum; they do not shift at fixed orbital energy since keplerian
energies and mean motions vary only with semi-major axis.  Thus for
eccentric planets, one must distinguish the rate of change of the
planetary angular momentum (torque) from the rate of change of orbital
energy.  Smoothing of the planetary potential along epicycles tends to
reduce the strength of the highest-$m$ Lindblad resonances that
dominate the torque on circular orbits.  All of these perturbative
effects are expected to become significant when $e\gtrsim h/r$, where
$h\equiv c_s/\Omega$ is the disk thickness.  Planets of terrestrial
mass achieved such eccentricities in N2005's simulations.  On the
other hand, at about the same eccentricity the epicyclic velocities
become supersonic, which will probably raise significant shocks, a
non-perturbative effect.  Thus both the sign and magnitude of the
changes in the mean torque due to turbulently excited eccentricity are
unclear.  We are not aware of any systematic investigation of this
issue, nor have we attempted one.  Throughout this paper, we treat the
planetary orbit as circular.

\item Even for circular orbits, stochastic changes in semi-major axis
on timescales long compared to the period but short compared to the
nominal migration time may significantly alter the mean orbital
lifetime.  This is particularly likely when the local mean torque
varies strongly with radius, as it does in the alpha disks studied by
\citet{mg04}.  Because of rapid changes in opacity in certain
temperature regimes, the disks exhibit sharp peaks in the local
migration timescale.  Under laminar conditions, orbital lifetimes can
be dominated by the time required to drift across these peaks.
Turbulent fluctuations in the torque, however, could allow planets to
``jump over'' the peaks, thus shortening the lifetime.  On the other
hand, diffusion to large radii where the local migration time is long
might prolong the orbital lifetime.

\item Finally, orbital diffusion should allow the occasional planet to
survive much longer than the mean lifetime defined by its birthplace.
If sufficiently many planets are formed, this might in principle help
to reconcile theoretical predictions of rapid \emph{mean} migration
with the observations at least of the longer-period planets (though it
seems unlikely to explain hot Jupiters).

\end{enumerate}

The methods adopted in this paper allow us to explore the second and
third issues above, but not the first.  In \S\ref{sec:FP}, we develop
a simple advection-diffusion or Fokker-Planck equation for the
probability distribution function of planetary orbital angular
momentum in the presence of torque fluctuations.  Parametrizations of
the advection and diffusion coefficients are calibrated against
semianalytic formalisms and the simulations of N2005, respectively.
\S3 explores analytic steady-state and time-dependent solutions to
this equation in steady disks whose properties are power laws in
radius.  \S4 and \S5 present numerical, time-dependent solutions in
such a disk (the Minimum Mass Solar Nebula) and in alpha disks with
realistic opacities.  A summary of our main results and conclusions is
given in \S6.

\section{Advection-diffusion Equation for Planetary Migration}\label{sec:FP}

Planets migrate as a result of torques exerted by the disk.  Because
of turbulence, the torque is a stochastic function with a mean part
$\bar\Gamma$ and a fluctuating part $\delta \Gamma$.  For a given
planetary mass ($M_p$) and a given time-averaged surface density and
thickness of the disk [$\bar\Sigma(r),\,\bar h(r)$ respectively], the
mean torque depends only on the semimajor axis, $r_p$, of the
planetary orbit, which is assumed to be circular.  The diffusion
equation is most conveniently derived in terms of the orbital angular
momentum\footnote{Throughout our analysis, we focus on the limit $M_p
\ll M_*$, where $M_*$ is the mass of the central star.}
$J=M_p\sqrt{GM_*r_p}$, rather than $r_p$, as the spatial independent
variable.

The fluctuating part of the torque depends on time as well as $J$:
$\delta\Gamma=\delta\Gamma(t,J)$.  By construction,
$\overline{\delta\Gamma}=0$.
The fluctuating torque is taken to be a temporally stationary stochastic
process with a finite amplitude and correlation time, so that the integral
\begin{equation}\label{Ddef}
D(J)\equiv\frac{1}{2}\int\limits_{-\infty}^\infty \overline{
\delta\Gamma(t-\,\tau/2,J)\delta\Gamma(t+\,\tau/2,J)}\,d\tau
\end{equation}
exists. $D(J)$ will play the role of diffusion coefficient.
The correlation time can then be defined by
\begin{equation}\label{tcorr}
\tau_{\rm c}\equiv \left.D(J)\right/\overline{\left[\delta\Gamma(t,J)\right]^2}\,.
\end{equation}
Temporal stationarity implies that $D(J)$,
$\overline{\left[\delta\Gamma(t,J)\right]^2}$, and $\tau_{\rm c}$
are independent of $t$.

The desired advection-diffusion equation derives from a Fokker-Planck
equation, and its formal derivation is completely standard
\citep[\emph{e.g.}][]{vanKampen}, but it is useful to repeat the
derivation here in order to emphasize the underlying assumptions and
to ask whether they are justified in this case.  Two important
assumptions have already been introduced: the finiteness of
Eqs.~(\ref{Ddef}) \& (\ref{tcorr}).

Let $P(\Delta J\,| \Delta t, J)$ be the probability that a planet of
initial angular momentum $J$ suffers a change $\Delta J$ during time
interval $\Delta t$.
Moments of the change are
\begin{equation}\label{moments}
\overline{(\Delta J)^n}\equiv
\int\limits_{-\infty}^\infty (\Delta J)^n
P(\Delta J\,|\Delta t,J)\,d\Delta J\,.
\end{equation}
Of course, except for $\overline{(\Delta J)^0}=1$, 
these depend implicitly on $J$ and $\Delta t$.

Next, let $f(t,J)$ be the probability that the planet has angular momentum
$J$ at $t>0$ given some initial condition at $t=0$.
The probability at $t+\Delta t$ is
\begin{eqnarray}\label{master}
f(t+\Delta t,J)&=&
\int\limits_{-\infty}^\infty P(\Delta J\,|\Delta t,J-\Delta J)
f(t,J-\Delta J)\,d\Delta J\nonumber\\
&=&\int\limits_{-\infty}^\infty
\sum\limits_{n=0}^\infty\frac{(-1)^n}{n!}(\Delta J)^n\frac{\partial^n}
{\partial J^n}\left[P(\Delta J\,|\Delta t,J)f(t,J)\right]\,d\Delta J
\nonumber\\
&=&\sum\limits_{n=0}^\infty\frac{(-1)^n}{n!}
\frac{\partial^n}{\partial J^n}\left\{\vphantom{\frac{a}{a}}
\overline{(\Delta J)^n}f(t,J)\right\}\,.
\end{eqnarray}
In the Fokker-Planck approximation, the changes are presumed small
compared to $J$ itself but cumulative, so that one can ignore the
terms $n>2$ in the sum above.  More will be said about the
justification for this below.

Clearly $\overline{(\Delta J)^0}=1$,
$\overline{\Delta J}=\bar\Gamma(J)\Delta t$, and
\begin{eqnarray*}
\overline{(\Delta J)^2}&=&
\overline{\left[\bar\Gamma(J)\Delta t~+~\int\limits_0^{\Delta t}
\delta\Gamma(t')\,dt'\right]^2}\\[1ex]
&=&\bar\Gamma(J)^2(\Delta t)^2~+~\int\limits_0^{\Delta t}\,d t_m
\int\limits_{-2 t_m}^{2 t_m}\,d \tau~
 \overline{\delta\Gamma( t_m-\,\tau/2,J)\delta\Gamma( t_m+\,\tau/2,J)}.
\end{eqnarray*}
The next important assumption is that $\tau_{\rm c}$ is short compared
to the timescale on which $J$ changes by order itself---let us write
$t_J$ for the latter timescale---so that we may take
$\Delta t$ in the range $\tau_{\rm c}\ll\Delta t\ll t_J$.  In
that case, the double integral above $\approx 2 D(J)\Delta t$.
At this point,
\[
f(t+\Delta t,J)\approx ~f(t,J)~+~\left[
-\frac{\partial}{\partial J}\left(\bar\Gamma f\right)~+
\frac{\partial^2}{\partial J^2}\left(D f\right)\right]\Delta t
~+~O\left[(\Delta t)^2\right],
\]
where the $O[(\Delta t)^2]$ part includes the $(\bar\Gamma\Delta t)^2$
contribution to $\overline{(\Delta J)^2}$.  Neglecting these terms on the
grounds that $\Delta t\ll t_J$, one has the desired
advection-diffusion equation:
\begin{eqnarray}\label{diffusion}
\frac{\partial f}{\partial t} +\frac{\partial F_J}{\partial J} &=&0
\nonumber\\[1ex]
F_J &\equiv& \bar\Gamma(J) f(t,J)~-~\frac{\partial }{\partial J}
\left[\vphantom{\frac{1}{2}}
D(J)f(t,J)\right]\,.
\end{eqnarray}

An interesting feature of the above result is that if the diffusion
coefficient $D(J)$ varies with $J$, then it influences the \emph{mean}
migration rate.  That is to say, one can rewrite the flux in
Eq.~(\ref{diffusion}) as
\[
F_J = \left(\bar\Gamma-\frac{\partial D}{\partial J}\right)f~-~
D\frac{\partial f}{\partial J}\,,
\]
so that $-\partial D/\partial J$  contributes to the advection term.

\subsection{Calibration of the diffusion and advection coefficients}
\label{subsec:calib}

We have tried to abstract a value and scaling for $D(J)$ from
the MHD simulations of N2005; the results of \citet{lsa04} were difficult
to translate into our framework.
Nelson reports that the typical timescale of fluctuations in $\delta\Gamma$
is of order half an orbital period.  Hence we take
\begin{equation}\label{tcorrsim}
\tau_c\approx\pi\left(\frac{r_p^3}{GM_*}\right)^{1/2}
=\frac{\pi (J/M_p)^3}{(GM_*)^2}\,.
\end{equation}

A direct estimate of the correlation function of the data suggests,
however, that correlations may persist to much longer than an orbital
period, and it is conceivable that the integral (\ref{Ddef}) for
$D(J)$ does not even exist (Nelson 2005, private communication).  If
the correlation time does not exist, then the methods of this paper
are inapplicable to migration.  Such long-term correlations imply a
``memory'' in the turbulence, presumably involving persistent
structures in addition to the planet of interest: for example,
long-lived vortices \citep[e.g.][]{fn05,jk05}, or of course other
planets.  Pending further numerical evidence, we adopt as a working
hypothesis that $\tau_{\rm c}$ \emph{does} exist but with uncertain
magnitude.

To complete the diffusion coefficient, we require a parametrization of
the variance of the fluctuating torque in terms of time-averaged disk
properties.  This entails some guesswork, as the simulations of N2005
and \citet{np04} have explored
only a limited range of disk models and planetary radii.  A natural
scale for the gravitational force exerted on the planet by the local
gas is $2\pi G\Sigma M_p$, where $\Sigma$ is the surface density of
the disk.  This is the force that the planet would feel if suspended
just above the disk.  The corresponding natural scale for the torque
is $2\pi G\Sigma M_p r_p$.  Therefore, we postulate that the variance of
the torque fluctuations is
\begin{equation}\label{amptorque}
\overline{\left[\delta\Gamma(t,J)\right]^2}\approx
\left(\mathcal{C}_D\times 2\pi G\bar\Sigma r_p M_p\right)^2,
\end{equation}
where $\mathcal{C}_D$ is a dimensionless coefficient depending upon
the strength of the turbulence.

Since the parametrization (\ref{amptorque}) underlies all of our
results, it is worth more discussion.  The turbulence is accompanied
by density perturbations of r.m.s. amplitude $\delta\rho_\lambda$ on
scale $\lambda$.  An individual perturbation of this scale exerts a
gravitational force per unit mass $\sim G\delta\rho_\lambda\lambda^3
d^{-2}$ at distance $d$ if $\lambda<h$, and $\sim G\delta\rho_\lambda
h \lambda^2 d^{-2}$ if $\lambda>h$.  Presuming that the forces from
different perturbations add in quadrature, the mean-square force is
dominated by the closest such perturbation, at $d\sim\lambda$, since
the number of $\lambda$-scale perturbations within distance $d$
increases more slowly with $d$ than the mean-square contribution from
individual perturbations decreases: $\propto (d/\lambda)^3$ for $d<h$
and $\propto (d/\lambda)^2$ for $d>h$, versus $d^{-4}$.  Thus, the
r.m.s. force due to scale $\lambda$ is $\delta f_\lambda\sim
G\delta\rho_\lambda\lambda$ for $\lambda<h$ and $\delta f_\lambda\sim
G\delta\rho_\lambda h\sim G\delta\Sigma_\lambda$ for $\lambda>h$.  The
contributions to $D(J)$ from $\lambda<h$ are probably unimportant
because of the weighting by $\lambda$ and because smaller scales are
likely to have shorter correlation times.  If the density power
spectrum were as ``blue'' as white noise,
$\delta\rho_\lambda\propto\lambda^{-3/2}$, and if the correlation time
$\tau_{\rm c,\lambda}\propto\lambda$, then there would be equal
contributions from all scales $<h$, with the result that the total
contribution from these scales would be larger than the contribution
from $\lambda\gtrsim h$ by a logarithmic factor.  In reality, the
density power spectrum is surely red, since local magnetorotational
(henceforth MRI) simulations find that most of the turbulent magnetic
and kinetic energy resides at scales $\gtrsim h$ \citep{hgb95,bnst95}.
Thus, the dominant contribution to the force comes from scales
$\gtrsim h$.  Since $\delta f_\lambda\sim G\delta\Sigma_\lambda$ in
this regime, the total r.m.s. force fluctuation is proportional to the
total r.m.s. surface-density fluctuation, $\delta\Sigma$.

It remains to discuss how $\delta\rho/\rho$ and hence
$\delta\Sigma/\Sigma$ may depend upon the strength of the turbulence.
Density fluctuations may arise from fluctuations in gas pressure or in
entropy.  Local MRI simulations indicate that
$\delta\rho/\rho\sim\delta P/P\sim \delta B^2/8\pi\rho c_s^2 \approx
2\alpha$ \citep{sano03}, as might be expected from equipartition,
since the internal energy per unit mass associated with a density
fluctuation is $c_s^2\delta\rho/\rho$.  In a statistically steady
accretion disk, the local thermal time $\sim(\alpha\Omega)^{-1}$, so
if the correlation time of mechanical fluctuations is
$\sim\Omega^{-1}$, entropy fluctuations due to dissipation are
expected to be at most $\propto\alpha$.  Entropy fluctuations may also
arise by advection of material in the presence of a background entropy
gradient, $\delta S\approx-\delta\mathbf{x\cdot\nabla}S$.  The
simulations cited above indicate that the turbulent Reynolds stress
scales with the magnetic stress so $\delta\mathbf{v}\sim
\alpha^{1/2}c_s$, and we assume that the largest displacements have
frequencies $\sim\Omega$, so the r.m.s. displacement
$\delta\mathbf{x}\sim\alpha^{1/2}h$ in any direction.  Vertical and
horizontal entropy gradients are likely to have scale lengths $\sim h$
and $\sim r$, respectively, which would seem to make vertical
displacements more important.  However, the azimuthal force, and hence
torque, exerted by a mass element changes only at second order as it
is displaced vertically through $\delta z\sim\alpha^{1/2}h\ll h$,
so vertical displacements contribute at $O(\alpha)$.  
Radial displacements contribute at first order,
$(\delta\Sigma/\Sigma)_{dS/dr}\sim \alpha^{1/2}h/r$.  The vertically
uniform disk of N2005 has a radial entropy gradient, since the
sound speed is constant but the density is not; however,
since N2005 obtains $\langle\alpha\rangle\approx 5\times10^{-3}$ (in
agreement with previous work), it happens that $\alpha^{1/2}\approx
0.07=h/r$ in these models, so that we cannot distinguish between
$\delta\Sigma/\Sigma\sim\alpha$ and
$\delta\Sigma/\Sigma\sim\alpha^{1/2}h/r$.

It would be useful to calculate torque fluctuations in a variety of
disk models with varying entropy gradients, thicknesses, and (if
possible) strengths of turbulence.  Lacking such information,
we provisionally assume $\delta\Sigma\sim f(\alpha)\Sigma$, where
$f$ is some function (probably linear).
This leads to the parametrization (\ref{amptorque}) if $\alpha$
is a universal constant.  As will be seen, diffusion then dominates
mean migration, \emph{i.e.} inward drift at the rate predicted for a laminar disk,
in the outer parts of plausible disk models.
If $\delta\Sigma$ is also proportional to $h/r$, 
diffusion becomes even more dominant in the outer parts of a flaring disk.

N2005 reports that the r.m.s.  fluctuating torque per unit planetary
mass is $1.5\pm0.5\times 10^{-4}$ in code units.  In the same units,
$2\pi G\Sigma r_p\approx 3.3\times 10^{-3}$ (Nelson 2005, private
communication); the latter is independent of radius because
$\Sigma\propto r^{-1}$.  Therefore,
\begin{equation}\label{torqnorm}
\mathcal{C}_D= 0.046
\end{equation}
is taken as the reference value of the dimensionless factor in equation
(\ref{amptorque}).

Following Eq.~(\ref{tcorr}), the diffusion coefficient
$D(J)$  is the product of the correlation time $\tau_{\rm c}$
(\ref{tcorrsim}) and the torque variance (\ref{amptorque}).
Both factors are uncertain, but within the framework of this
advection-diffusion approach, they combine into a single parameter.
We hope that future numerical simulations will refine our estimates
of this parameter and, more importantly, test the validity of
the scalings (\ref{tcorrsim}) and (\ref{amptorque}).

For comparison, \citet{w97a}'s expression for the mean Lindblad torque
density,
\begin{equation}\label{eq:ward}
\left(\frac{d\Gamma}{dr}\right)_{LR}=\mbox{sign}(r-r_p) \frac{2
\mu^2\bar\Sigma r_p^4\Omega_p^4} {r (1+4\xi^2)\kappa^2}\,m^4\,\psi^2,
\end{equation}
leads to the scaling
\begin{equation}\label{eq:meantorque}
\bar\Gamma\propto \frac{GM_p^2r_p^3\bar\Sigma}{M_* h^2}.
\end{equation}
For consistency with our calibration of torque fluctuations, we also
calibrate this relation against Nelson's simulations of laminar disks,
in which the type I torque per unit mass is $\simeq
1.5\times10^{-6}M_p/M_\oplus$ in code units.  The constant of
proportionality in Eq.~(\ref{eq:meantorque}) is then $\simeq -4.8$.  Using
Eq.~(\ref{eq:meantorque}) implies that the mean torque is always
negative.  This is fine for plausible power-law disks but might be
misleading for alpha disks, where the surface density and temperature
can be highly structured, allowing in principle for positive
(or at least much reduced) $\bar\Gamma$ at some radii.  Thus
when modeling alpha disks we use Eq.~(\ref{eq:ward}) directly, as
described in \citet{mg04}.

\section{Analytic results}\label{sec:analytic}

Power-law disk models such as the Minimum-Mass Solar Nebula (hereafter
MMSN; \citet{ha81}) allow the advection-diffusion equation to be
solved analytically, at least in steady state.  Some aspects of
time-dependent evolution can also be determined analytically for this
class of disk models. Throughout this section, all disk models are
considered to be infinite in radius as well as steady.  While
unrealistic, this is technically convenient, and we are still able to
identify results that are likely to be sensitive to the details of the
outer boundary conditions: for example, the mean orbital lifetime.
For reference, the MMSN model adopted here has
mean surface density $\bar{\Sigma}\propto r^{-3/2}$ and aspect ratio
$h/r\propto r^{1/4}$.  

\subsection{Steady state}\label{subsec:steady}

A steady-state distribution of planets, $f(J)$, satisfies 
\begin{equation}\label{eq:steadystate}
\frac{\partial F_J}{\partial J}=S(J),
\end{equation}
where the right-hand side is a source term representing the rate of
formation of planets at radius $r_p$ with angular momentum
$J=M_p\sqrt{GM_*r_p}$.  Analytic solutions to Eq.~(\ref{eq:steadystate})
exist when mean torque and the diffusion coefficient are both
power-laws in $J$:
\begin{equation}
  \label{eq:powers}
\bar{\Gamma}(J)\propto J^\alpha,\qquad 
D(J)\propto J^\beta,
\end{equation}
and the source term is a Dirac delta function
\begin{equation}\label{eq:sourceterm}
S(J)=\delta(J-J_S)J_S\Lambda,
\end{equation}
where $\Lambda$ is the rate of planet formation at $J_S$.  We see
immediately that the flux of planets interior $(F_J^-)$ and exterior
$(F_J^+)$ to the source are both constant and related to each other by
\begin{equation}\label{sourceflux}
F_J^+ - F_J^- = J_S\Lambda.
\end{equation}
First we consider the limiting behavior as $J\rightarrow\infty$.  The
local timescale for secular inward migration is given by
$t_{\rm{mig}}= J/|\bar{\Gamma}(J)|$, and the local diffusion timescale
is given by $t_{\rm{diff}}=J^2/D(J)$.  The ratio, $t_{\rm diff}/t_{\rm
  mig}\propto J^{\alpha-\beta+1}$ indicates that the mean torque
becomes asymptotically negligible if $\alpha-\beta+1<0$.  (This
condition holds for the minimum-mass solar nebula, in which
$\bar{\Gamma}\propto -J^{-2}$ and $D\propto J$.)  For such disks the
solution to Eq.~(\ref{eq:steadystate}) at large $J$ exterior to the source
becomes $f(J)\approx (\mathcal{C}-JF_J^+)/D(J)$, where $\mathcal{C}$
is a constant of integration.  We insist that $F_J^+\ge0$ since there
should not be an incoming flux of planets from infinity.  Thus the
numerator eventually becomes negative. The diffusion coefficient
$D(J)$ is unquestionably positive, so $f$ becomes negative at
sufficiently large $J$.  But this is absurd, since $f(J)dJ$ is a
probability.  Thus we cannot have both a positive flux and a positive
$f$ at large $J$ in steady-state, and it follows that $F_J^+=0$ and
$F_J^-=-J_S\Lambda$.  All the flux generated at the source is accreted
through the inner edge of the disk.

We now discuss Eq.~(\ref{eq:steadystate}) in full, emphasizing
behavior at small and large $J$.  Let
\begin{equation}\label{eq:mu}
\frac{d\mu}{dJ}=-\frac{\bar\Gamma(J)}{D(J)}\,,
\end{equation}
so that $\exp[\mu(J)]$ is an integrating factor for Eq.~(\ref{eq:steadystate}).
Then
\begin{equation}\label{eq:fss}
f^{\pm}(J)= \frac{1}{D(J)}e^{-\mu(J)}\left[e^{\mu(J_S)}D(J_S) f(J_S)- F_J^{\pm}
\int\limits_{J_S}^J e^{\mu(\bar J)}d\bar J\right].
\end{equation}
As before, the upper (lower) sign on $f^{\pm}$ and $F_J^{\pm}$ applies
to $J>J_S$ ($J<J_S$).  When $\bar\Gamma(J)$ and $D(J)$ are power laws,
$f^\pm$ can be evaluated explicitly, but we will not write it
out. Normally the
mean torque $\bar\Gamma<0$, so that $d \mu/dJ >0$.  Then we can choose
the constant of integration in Eq.~(\ref{eq:mu}) so that $\mu\le0$ for
$J\le J_S$ and $\mu\ge0$ for $J\ge J_S$.  All disks of interest in
this paper are dominated by the mean torque at small radii and by
diffusion at large $J$.  Hence $\exp[-\mu(J)]$ becomes large at small
$J$, and so for a well-behaved solution, the contents of the square
brackets in Eq.~(\ref{eq:fss}) must tend to zero as $J\to0$.  This
defines a relationship between $f(J_S)$ and $F_J^{-}$.  For $J>J_S$,
$e^{\mu(J)}\ge 1$, so the integral increases at least as fast as
$J-J_S$.  As before, if the disk is indefinitely extended then
we must have $F^+_J=0$ to prevent $f^+(J)$ from becoming negative at
large $J$.

The main conclusion of this discussion is that {\it as long as
the mean torque is nonpositive in the outer disk, the flux of
planets diffusing to infinite radius vanishes.}
We have reached this conclusion by considering steady solutions but
will assume that it applies generally.

\subsection{Time-dependent evolution}\label{subsec:timedep}

The time-dependent problem is more difficult, but we
can determine some general characteristics of the solution
by analytic means, most importantly that {\it the probability of
surviving without accretion in an infinite steady disk, 
though declining monotonically with time, can
have a long power law tail}.  For the special case of the MMSN, where
$D\propto J$ and $\bar\Gamma\propto-J^{-2}$, this probability declines
only as $t^{-1}$.   The asymptotic form of the solution at late
times is self-similar except near $J=0$ and is described explicitly in
dimensionless units by equation (\ref{eq:ssim}).

The evolution of a planet created with $J=J_S$ at $t=0$ can be represented
by solving Eq.~(\ref{diffusion}) with a source term $\delta(t)\delta(J-J_S)$.
The Laplace transform
\begin{equation}
  \label{eq:laplace}
  \hat f(s,J)\equiv\int\limits_0^\infty e^{-st} f(t,J)\,dt\,,
\end{equation}
reduces the time-dependent problem to
\begin{equation}
  \label{eq:odes}
\frac{\partial^2}{\partial J^2}\left[D(J)\hat f(s,J)\right]-
\frac{\partial}{\partial J}\left[\bar\Gamma(J)\hat f(s,J)\right]
-s\hat f(s,J)=-\delta(J-J_S).
\end{equation}
Unfortunately, although Eq.~(\ref{eq:odes}) is just an ordinary
differential equation in $J$, we cannot solve it completely even for
power law disks described by Eq.~(\ref{eq:powers}).  The difficulty is
in part that physically interesting cases have
$\alpha-\beta<-1$,
which is precisely the condition for Eq.~(\ref{eq:odes}) to have an
irregular singular point at $J=0$.
However, a formal solution can be obtained in powers of $s$,
\begin{equation}
  \label{eq:expans}
  \hat f(s,J)=\sum\limits_{n=0}^{\infty}\frac{(-s)^n}{n!}f_n(J).
\end{equation}
Replacing $\exp(-st)$ by its Taylor series
in the Laplace transform (\ref{eq:laplace})
shows that $f_n(J)$ is the $n^{\rm th}$ moment of $f(t,J)$ with respect
to $t$,
\begin{equation}
  \label{eq:moment}
  f_n(J)=\int\limits_0^\infty t^n f(t,J)\,dt.
\end{equation}
More interesting than the moments of $f$ are the moments of
the corresponding flux:
\begin{equation}
  \label{eq:momJ}
  F_n(J)\equiv -(Df_n)'+\bar\Gamma f_n=
\int\limits_0^\infty t^n F_J(t,J)\,dt.
\end{equation}
The probability that the planet survives longer than $t$ before
accretion, $P(t)$, is related to the flux at the origin by
$dP/dt=F_J(t,0)$.  Therefore,
\begin{equation}\label{eq:tmom}
  -F_n(0)=-\int\limits_0^\infty t^n\frac{dP}{dt}\,dt
\end{equation}
is the $n^{\rm th}$ moment of the lifetime, and in particular,
$-F_1(0)$ is the mean lifetime.  

Thus we are motivated to find the $f_n(J)$.  Substituting
Eq.~(\ref{eq:expans}) into Eq.~(\ref{eq:odes}) leads to
\begin{eqnarray}
  \label{eq:f0}
  [D(J)f_0(J)]''-[\bar\Gamma(J)f_0(J)]'&=&-\delta(J-J_S), \\
  \label{eq:recur}
  [D(J)f_n(J)]''-[\bar\Gamma(J)f_n(J)]'&=&-n f_{n-1}(J)
\qquad n>0,
\end{eqnarray}
where the primes indicate derivatives with respect to $J$.  The
solution to Eq.~(\ref{eq:f0}), which is formally identical to the
steady-state case (\ref{eq:steadystate}), is a Green's function for
eq.~(\ref{eq:recur}).  It can be constructed from
homogeneous solutions $y_0(J)$ and $y_1(J)$ with fluxes $0$ and $-1$,
respectively:
\begin{eqnarray}
  \label{eq:y0}
[D(J)y_0(J)]'~-\bar\Gamma(J)y_0(J) &=& 0,\\
  \label{eq:y1}
[D(J)y_1(J)]'~-\bar\Gamma(J)y_1(J) &=& 1.
\end{eqnarray}
The Green's function is
\begin{equation}
  \label{eq:green}
  G(J,J_S) = y_1(J_<)y_0(J_>)/y_0(J_S),\qquad J_<\equiv\min(J,J_S),\qquad
J_>\equiv\max(J,J_S).
\end{equation}
The formal solution to the system (\ref{eq:f0})-(\ref{eq:recur}) is then
\begin{eqnarray}
  \label{eq:ssol}
  f_0(J) &=& G(J,J_S),\nonumber\\
f_n(J) &=& n!\int\limits_0^\infty dJ_1
\ldots\int\limits_0^\infty dJ_n\,G(J,J_1)G(J_1,J_2)\ldots G(J_n,J_S).
\end{eqnarray}
The existence of the moments (\ref{eq:moment}) therefore
depends upon the convergence of the integrals in (\ref{eq:ssol}).

An important special case is the MMSN, where $D(J)=D_0 J$ and
$\bar\Gamma=-\Gamma_0 J^{-2}$.  It is convenient to choose
$J_*\equiv(\Gamma_0/D_0)^{1/2}$ as the unit of angular momentum and
$t_*\equiv J_*/D_0$ as the unit of time so that $D(J)\to J$ and
$\bar\Gamma(J)\to -J^{-2}$.  The homogeneous solutions (\ref{eq:y0})
become
\begin{eqnarray}
  \label{eq:y0MMSN}
y_0(J) &=& J^{-1}e^{x^2}\qquad\mbox{where}~ x\equiv\frac{1}{J\sqrt{2}}\,,\\
\label{eq:y1MMSN}
y_1(J) &=& 1 - \pi^{1/2} xe^{-x^2}\mbox{erfc}(x)
\nonumber\\[2ex]
&\approx&
\cases{
   J^2-(3!!)J^4+(5!!)J^6-\ldots & $J\to0^+$ \cr
   1-\sqrt{\frac{\pi}{2}}\sum\limits_{n=0}^\infty\frac{1}{(2n)!!}J^{-2n-1}
~+\sum\limits_{n=1}^\infty\frac{1}{(2n-1)!!}J^{-2n} & $J\to\infty$.
}
\end{eqnarray}
The first expansion is only asymptotic, in keeping with the fact that $J=0$
is an irregular singular point of Eq.~(\ref{eq:y1}), 
whereas the second converges for all $J\ne0$.

Armed with these results, we can now determine the convergence of the
integrals (\ref{eq:ssol}).  Because $y_0(J)\propto J^{-1}$ as
$J\to\infty$, it follows from (\ref{eq:green}) that the integrand for
$f_1(J)$ is proportional to $J_1^{-1}$ as $J_1\to\infty$.  Thus the
integral over $J_1$ is logarithmically divergent.  Consequently
{\it the mean lifetime defined by Eq.~(\ref{eq:tmom}) is not finite}.
Since the divergence is only logarithmic, one may guess (correctly)
that $P(t)\propto t^{-1}$ at late times in the infinite MMSN.

The long tail in the distribution of lifetimes results from diffusion
to large radii where torques are weak and planets linger.  
It is natural to seek a self-similar solution to Eq.~(\ref{diffusion}) that
reflects this behavior.  Unfortunately $J_*$ defines a 
preferred scale that prevents self-similarity.
Omitting $\bar\Gamma$ from Eq.~(\ref{diffusion}) on the grounds that
this term is unimportant at $J\gg J_*$ removes the obstruction:
\begin{equation}
  \label{eq:nogamma}
  \frac{\partial \tilde f}{\partial t}(t,J) =
\frac{\partial^2}{\partial J^2}[J \tilde f(t,J)],
\end{equation}
in the present dimensionless units with $D(J)\propto J$ as in the
MMSN.  The tilde distinguishes the solution to the approximate
equation (\ref{eq:nogamma}) from the exact solution $f(t,J)$ of
Eq.~(\ref{diffusion}). We require $\tilde f\to0$ as $J\to\infty$.  On
the other hand, $\tilde f(t,J)$ should be positive and finite at $J=0$
for the following reason. At late times such that $J\ll\sqrt{D(J)t}$,
we expect $f(t,J)$ to match onto a multiple of the constant-flux
steady-state solution defined by Eq.~(\ref{eq:y1MMSN}): $f(t,J)\approx
-F_J(t,0)y_1(J)$.  We may replace $f$ with $\tilde f$ in this relation
if $J\gg J_*$, and in that region $y_1(J)\approx 1$.
Eq.~(\ref{eq:nogamma}) can be solved with these boundary conditions by
separation of variables.\footnote{This is eased by the change of
variables $x\equiv\sqrt{2J}$ and $g(x)\equiv x\tilde f(J)$, which
produces the Bessel equation of order one on the righthand side.
Hankel transforms can then be used.}
For arbitrary positive initial conditions,
\begin{equation}
  \label{eq:ssim}
  \tilde f(t,J)\propto t^{-2}e^{-J/t}\qquad\mbox{as}~t\to\infty.
\end{equation}
As promised, this has self-similar form: $J$ enters only in the
combination $J/t$ [or $J/(D_0t)$ in dimensionful units].  It can be
verified that Eq.~(\ref{eq:ssim}) is an exact solution of
Eq.~(\ref{eq:nogamma}), and its integral
over all $J$ is $\propto t^{-1}$, in agreement with
our expectation for $P(t)$ at late times.

For other values of the indices $\alpha$ \& $\beta$ satisfying
$\alpha<\beta-1$, we have the following situation.  Again,
$\bar\Gamma$ is negligible at large $J$, so Eqs.~(\ref{eq:y0}) \&
(\ref{eq:green}) yield $G(J_1,J_S)\propto 1/D(J_1)$ at $J_1\gg J_S$.
Therefore the integral (\ref{eq:ssol}) for the first moment $f_1(J)$
diverges if $\beta\le1$, and consequently, the mean lifetime converges
only if $\beta>1$.  In other words, the mean lifetime exists
only if $t_{\rm diff}$ increases more slowly than $r^{1/2}$.

\subsubsection{A useful theorem about the eigenvalue spectrum}
\label{subsubsec:eig}

The following technical result helps to
interpret differences between the late-time behaviors of the planetary
distribution in the MMSN and in alpha disks.

The large-$t$ behavior of $f(t,J)$ is dominated by the singularities
of $\hat f(s,J)$ at largest $\mbox{Real}(s)$.  These correspond to
eigenvalues $\{\lambda\}$ obtained by separating variables in
Eq.~(\ref{diffusion}) with assumed time dependence $\exp(-\lambda t)$.
If the eigenvalues are real and discrete,
$0<\lambda_1<\lambda_2<\ldots$, then $f(t,J)\propto
f_{\lambda_1}(J)\exp(-\lambda_1 t)$ at late times, where
$f_{\lambda}(J)$ is the eigenfunction corresponding to $\lambda$.  (We
assume that these modes are complete.)  On the other hand, if the
eigenvalue spectrum is continuous and extends to arbitrarily small
positive $\lambda$, then the late-time decay is slower than
exponential, perhaps a power law. Defining
\[
g_\lambda(J)\equiv\int\limits_J^\infty f_\lambda(\bar J)d\bar J,
\]
and setting $w\equiv e^\mu$ and $p\equiv e^\mu D$, with $\mu$ as in
Eq.~(\ref{eq:mu}),
one can re-cast the separated version of Eq.~(\ref{diffusion}) in
self-adjoint Sturm-Liouville form
\begin{equation}\label{eq:SL}
\frac{d}{dJ}\left[p(J)\frac{d}{dJ}g_\lambda(J)\right]~+\lambda w(J)
g_\lambda(J)=0.
\end{equation}
In all disks of interest to us, $f\to0$ as $J\to0$ and the integral of
$f$ over $J$ is finite, whence $g'_\lambda(0)=0$ and $g_\lambda(\infty)=0$.
A theorem of \citet{dk05} then asserts that if the limit
\begin{equation}\label{eq:dk}
\lim_{J\to\infty}\left[\int_0^{J} w(x)dx\right]^{1/2}
\left[\int_{J}^\infty \frac{dx}{p(x)}\right]^{1/2}
\end{equation}
vanishes, then the spectrum is discrete and positive.  
[The theorem also requires $p$ and $w$ to be positive and
continuous, which is always true for us, and
it encompasses ``quasilinear'' generalizations of Eq.~(\ref{eq:SL}).]
It is easy to show that in disks where $t_{\rm mig}/t_{\rm diff}$ 
tends to zero as $J\to0$ and to infinity as $J\to\infty$, the limit
(\ref{eq:dk}) is equivalent to
\begin{equation}\label{eq:lim}
\lim_{J\to\infty}\left[J\int_J^\infty\frac{d\bar J}{D(\bar J)}\right]^{1/2}.
\end{equation}
In the MMSN [$D(J)\propto J$], this limit diverges, so a continuous
spectrum extending to $\lambda=0^+$ is allowed and indeed must exist,
or else we would not have the self-similar behavior discussed
above.  In the alpha disks studied numerically in
\S\ref{sec:numerical}, on the other hand, it appears that $t_{\rm
diff}\equiv J^2/D(J)\to0$ as $J\to\infty$ [Fig.~(\ref{fig:4dprop})],
so the limit vanishes and a discrete spectrum is implied.  This is
clearly true of power-law disks with $\alpha-\beta+1<0$ and $\beta>2$,
so the survival probability $P(t)$ should decay exponentially in these
disks.  We conjecture that the continuum exists and $P(t)$ follows a
power law for all $\beta<2$, because further consideration of
Eqs.~(\ref{eq:ssol}) indicates that the $n^{\rm th}$ moment of the
lifetime exists only if $\beta>2n/(n+1)$.

\section{Numerical method and tests}\label{sec:numerical}

The diffusion equation (\ref{diffusion}) with a source term reads:
\begin{equation}\label{FPsource}
\frac{\partial F_J}{\partial J}=S(t,J)-\frac{\partial f}{\partial t}.
\end{equation}
We find numerical solutions with an implementation of the generic
relaxation code described in \cite{ham98}.  The finite-difference
scheme is fully implicit in time. Given a distribution $f(t,J)$ at
time $t$, we only need to solve the one-dimensional boundary-value
problem in $J$ to find the solution at time $t+dt$.  We thus relax
successively from one solution to another over a variable timestep,
$dt$.  Controls over the size of the timestep ensure that, at each
grid point, the change in each variable over $dt$ is within an
acceptable tolerance.  The system is described by two first-order
differential equations equivalent to Eq.~(\ref{FPsource}), supplemented
with a third differential equation that defines the numerical domain.
For simplicity, we do not implement an adaptive mesh but use instead a
fixed grid on the numerical domain, chosen to be uniformly spaced in
$\log r$.  We take as variables the set
$(f,\partial(Df)/\partial J,J)$. Numerical stability of the solutions is
achieved with centered-differencing schemes for $f$ and $J$, and an
upwind-differencing scheme for $\partial(Df)/\partial J$.

Three boundary conditions are required.  The numerical domain extends
from the inner edge of the disk, at $ \simeq 10^{-2}\au$, out to
$1000\au$.  At the inner boundary we trivially set the value of $J$
and are free to fix $f=0$.  At the outer boundary we demand the flux
be zero.  In steady-state, as discussed in \S\ref{subsec:steady}, this
is a boundary condition that must be satisfied if diffusion dominates
at large radii.  To avoid potential complications near the disk edge,
we extend our numerical domain to sufficiently large radii that it
does not influence our results.

We have tested our numerical results for power-law models against the
analytic steady-state solutions described in \S\ref{subsec:steady}.
The numerical solutions converge satisfactorily for a modest grid
spacing (typically 1000 grid points over five decades in radius).
Additional tests at ten times this resolution show clear numerical
convergence of the steady-state results. Agreement with the late-time
behavior expected from the time-dependent self-similar solutions
described in \S\ref{subsec:timedep} will also be demonstrated below
(see Fig.~\ref{fig:MMSNsv}).

\section{Numerical Solutions}\label{sec:numres}

We study the combined effects of migration and diffusion on the
orbital evolution of low-mass proto-planets for two separate classes
of disk models: the Minimum Mass Solar Nebula (MMSN) and T Tauri alpha
disks. The MMSN model adopted here has a mean surface density
$\bar{\Sigma}\propto r^{-3/2}$ and an aspect ratio $h/r\propto
r^{1/4}$, which are normalized to $4200\,{\rm g\,cm^{-2}}$ and 0.1 at
$r\!=\!1\au$, respectively.  The T Tauri alpha disk model adopted is
identical to the irradiated disk described by \citet[][see their
Fig.~1]{mg04}, accreting steadily at a rate
$\dot{M}=10^{-8}M_\odot\yr^{-1}$.  We take as our reference model the
alpha disk with viscosity parameter $\alpha=0.02$.  For comparison, we
also consider a more massive and cooler alpha disk, with
$\alpha=10^{-3}$.  The inner edge of the disks is located at $0.01\au$
and $0.02\au$ in the MMSN and the alpha disks, respectively.  The mass
of the primary is $1{M_\odot}$ in the MMSN and $0.5{M_\odot}$ in the alpha
disks. These slight model differences have very little influence on
our results. Note that, while the disk models used are assumed to be
steady, we are studying the time-dependent orbital evolution of
planets embedded in these disks with equation (\ref{diffusion}). As a
consequence, our method can only apply to proto-planets of small
enough mass that they exert a negligible feedback on the disk
structure \citep[broadly speaking, $M_p\lesssim 10 M_\oplus$;
e.g.][]{mg04}.

Turbulence is assumed to be present everywhere in our disk models.
The magnitude of the corresponding diffusion coefficient is uncertain
through both the amplitude and the correlation time of the turbulent
torque fluctuations.  We quantify these uncertainties with a single
parameter, $\epsilon$.  If we define
\begin{equation}
\epsilon\equiv
\frac{\tau_c}{\tau_{\rm{orb}}}\left(\frac{\mathcal{C}_D}{0.046}\right)^2,
\end{equation}
then
\begin{equation}\label{eq:Deps}
D(J)=\tau_c\times\overline{\left[\delta\Gamma(t,J)\right]^2}=\epsilon\,(2\pi)^3(0.046)^2\frac{\bar{\Sigma}^2J^7}{G^2M_*^4M_p^5}.
\end{equation}
The calibration performed in \S\ref{subsec:calib} suggests that
$\epsilon\simeq 0.5$ is a reasonable value, which we have assumed to
be independent of radius. However, we also investigate the effects of
varying this parameter by several orders of magnitude.

\subsection{General Scalings}

It is useful to discuss a few general scalings specific to our disk
models before we discuss the details of time-dependent numerical
solutions. To quantify the relative importance of diffusion
\emph{vs.} migration in the models, we define a local diffusion time
$
t_{\rm{diff}}=J^2/D\propto \bar{\Sigma}^{-2}J^{-5}\propto
J^{-4k-5}
$
(for $\bar{\Sigma}\propto r^k$) and a local migration time,
$
t_{\rm{mig}}=J/|\bar\Gamma|\propto
h^2\bar{\Sigma}^{-1}J^{-5}M_p^{-1},
$
according to Eq.~(\ref{eq:meantorque}), so that $t_{\rm{mig}} \propto
J^3M_p^{-1}$ in the MMSN.  The ratio of these timescales has a particularly
simple dependence upon disk and planetary properties:
\begin{equation}\label{eq:migvsdiff}
\frac{t_{\rm mig}}{t_{\rm diff}}\propto \frac{h^2\bar\Sigma}{M_p}\,.
\end{equation}

One expects the evolution of planets located in
regions of a disk with $t_{\rm{mig}} \ll t_{\rm{diff}}$ to be
dominated by migration and vice-versa.  While the diffusion time is
independent of planet mass, the migration time is inversely
proportional to $M_p$. Therefore, lower mass proto-planets will be
more sensitive to the effects of turbulent torque fluctuations. In
addition, the two timescales differ in their dependence on the mean
surface density, $\bar \Sigma$, which tends to make planets in denser
disks more susceptible to the effects of diffusion. As we shall now
see, a general consequence of these scalings is that one expects the
effects of diffusion to dominate only in the outer regions of
proto-planetary disks (at least for the classes of disk models
considered here).

\begin{figure}
\plotone{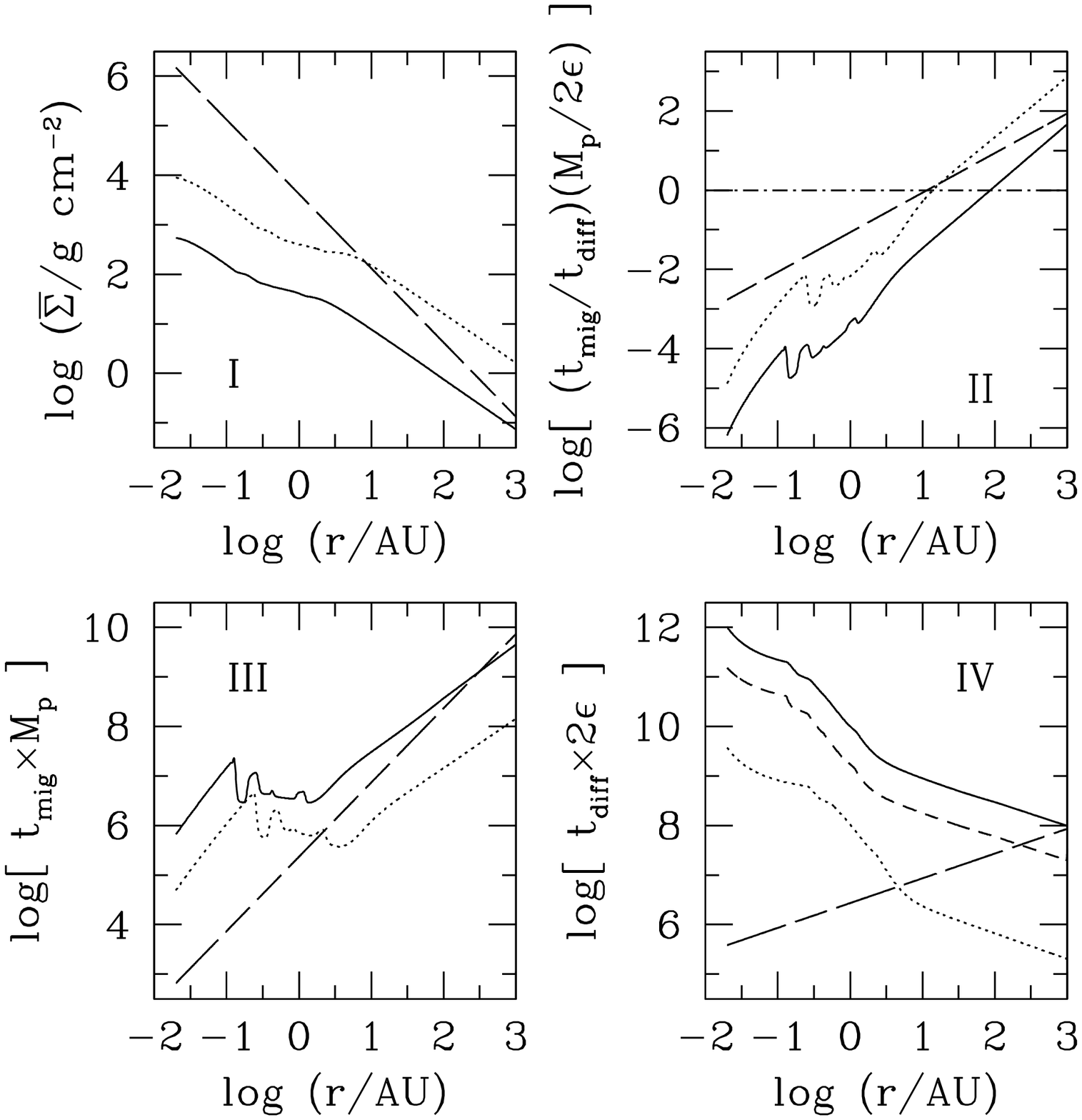}
\caption{In all four panels, quantities were plotted assuming a planet
mass $M_p=1 M_\oplus$ and torque fluctuations normalized to
$\epsilon=0.5$. The long-dashed lines correspond to the MMSN, while
the solid and dotted lines correspond to the $\alpha=0.02$ and
$\alpha=10^{-3}$ T Tauri alpha-disk models, respectively.  Panel I
shows mean surface density profiles.  Panel II shows ratio of local
migration time to local diffusion time in the three models.  The
horizontal dash-dotted line indicates the formal transition between
migration-dominated and diffusion-dominated regime for Earth-mass
planets.  Panels III and IV show local migration times and local
diffusion times, respectively.  The short-dashed line in Panel IV
corresponds to the local advection timescale due to inhomogeneous
diffusion, $J/(\partial D/\partial J)$, for the $\alpha=0.02$
alpha-disk model.\label{fig:4dprop}}
\end{figure}

Figure~\ref{fig:4dprop} shows profiles of various important quantities
in the MMSN and our two alpha disk models.  Panel I shows profiles of
mean surface density, $\bar\Sigma$.  Both the slope and normalization
of $\bar\Sigma$ profiles matter for the relative importance of
migration \emph{vs.} diffusion. Panels III and IV show profiles of local
migration and diffusion times in the various disk models. While
$t_{\rm{diff}}$ increases linearly with $J$ ($\propto r^{1/2}$) in the
MMSN, it decreases with increasing radius in the alpha disk models.
This is a consequence of the steeper $\bar\Sigma$ profile in the MMSN.
The profiles of $t_{\rm{mig}}$, derived from Eq.~(\ref{eq:meantorque})
for the MMSN and from the detailed calculations\footnote{Specifically,
  we use the 3D model in which the gas is assumed to be vertically
  extended over a full disk scale height for the torque calculation.}
of \citet{mg04} for the alpha disk models, show local migration times
which decrease toward small radii in both classes of models. Finally,
panel II shows the ratio of timescales, $t_{\rm{mig}}/t_{\rm{diff}}$,
which is important to characterize the behavior of planets embedded in
these disks. As mentioned earlier, in all three models, this ratio
grows above unity (indicated by the horizontal dash-dotted line in
panel II) only at radii $\gtrsim 10 \au$ if $M_p\lesssim 1\,M_\oplus$,
indicating that diffusion should dominate over migration only in the
outer regions of these disks (and this despite the increasing value of
$t_{\rm{diff}}$ with radius in the MMSN).

Although very useful, the profiles shown
in Figure~\ref{fig:4dprop} do not capture all of the complexity of the
migration-diffusion problem. First, all the quantities plotted in
panels II-III-IV were derived for a planet mass $M_p=1 M_\oplus$ and a
torque fluctuation normalization $\epsilon=0.5$. The profiles are
easily rescaled for different values of these parameters (as indicated
by labels in Fig.~\ref{fig:4dprop}), but it is clear that the exact
location of the transition from migration- to diffusion-dominated
evolution depends on the proto-planetary disk model adopted, the value
of the planet mass, and the normalization $\epsilon$ of the diffusion
coefficient.  Nevertheless, 
the dominance of diffusion over migration in the outer
regions of these disks is robust because
$\bar\Sigma h^2$ tends to increase with radius in disk models of
interest [Eq.~(\ref{eq:migvsdiff})].

Secondly, as shown in panel III, in a disk model with realistic
opacities, sudden and non-monotonic variations in the value of
$t_{\rm{mig}}$ at radii $\lesssim 3\au$ could complicate the orbital
evolution of planets substantially, by stalling their migration as
they reach the disk inner regions. This is best explored with global
numerical solutions of the time-dependent Fokker-Planck equation, as
seen in \S\ref{subsec:burst}.

A third complication arises from the radial dependence of
the diffusion driven by turbulent torque fluctuations. As mentioned in
\S\ref{sec:FP}, the radial variation of the diffusion coefficient
leads to an additional contribution to the advected flux of
planets. Since $-\partial D/\partial J\propto -(7+4k)\,J^{6+4k}$
for $\bar{\Sigma}\propto r^k$,
and since $k=-3/2$ in the MMSN and $k>-3/2$ in the alpha disk models,
this additional term tends to advect planets inward in all the disk
models considered.  We can define a local advection timescale
associated with this new contribution, $J/(\partial D/\partial J)$,
and compare it to the values of $t_{\rm{mig}}$ and $t_{\rm{diff}}$
defined above. We find that this advection timescale is everywhere
exactly equal to $t_{\rm{diff}}$ in the MMSN, while it is shorter than
$t_{\rm{diff}}$ everywhere in our alpha disk models (as shown
explicitly for the $\alpha=0.02$ case by the short-dashed line in
panel IV of Fig.~\ref{fig:4dprop}). Consequently, even in disk regions
where diffusion nominally dominates over migration in our models (as
measured by the ratio $t_{\rm{diff}}/t_{\rm{mig}}$), this additional
contribution due to radially inhomogeneous diffusion should still
cause a rather significant effective advection of planets towards the
star.

\subsection{Continuous and Burst Formation of Planets}\label{subsec:burst}

We have studied two different types of time-dependent models for the
diffusive migration of planets in proto-planetary disks, corresponding
approximately to a continuous and a burst scenario for planet
formation. In both cases, for the sake of clarity, the source of
planets was localized, taking the form of a sharply peaked Gaussian
approximating the delta source function of
Eq.~(\ref{eq:sourceterm}). We have verified that our results do not
depend sensitively on the detailed shape of the Gaussian source
function adopted. As we shall see, the predictions for these rather
different scenarios for planet formation are broadly consistent with
each other.

\begin{figure}
\plotone{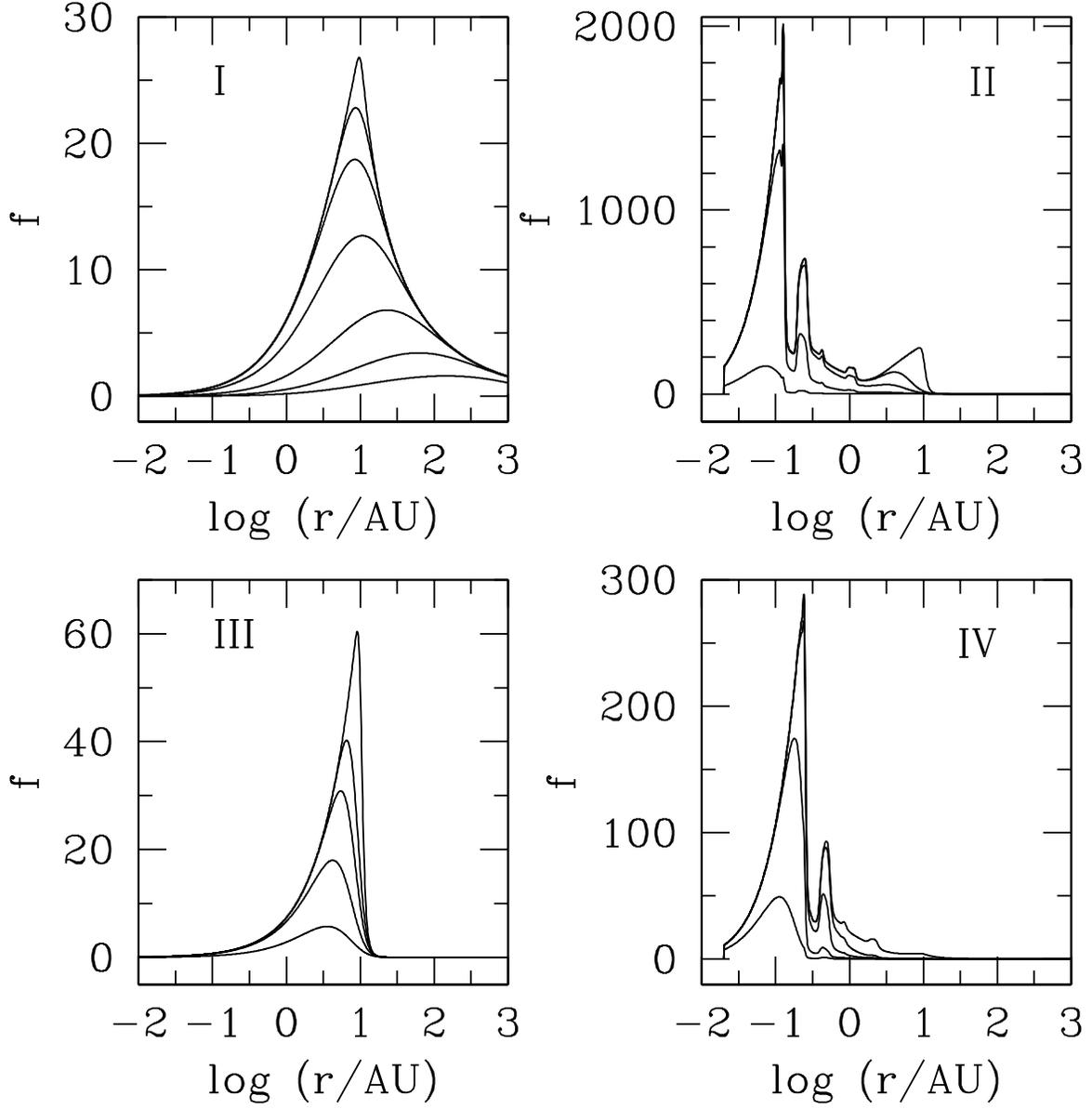}
\caption{Snapshots of the decay of steady-state distributions after
source shut-off.  In each case, the source formed Earth-mass planets
at $r_S=10\au$ at a rate $\Lambda=(10^5\yr)^{-1}$ until shut-off.  The
disk model in Panels I and III is the MMSN with torque fluctuations
normalized to $\epsilon=0.5$ and $0.02$, respectively.  The disk
models in Panels II and IV are alpha disks with viscosity parameters
$\alpha=0.02$ and $10^{-3}$, respectively ($\epsilon=0.5$).  Snapshots
are shown at the following times after source shut-off, in log-years:
(I) 5,5.5,6,6.5,7,7.5; (II) 6.8,7,7.2,7.4; (III) 5.8,6,6.2,6.4; (IV)
5.8,6,6.2,6.4.\label{fig:4steady}}
\end{figure}

Figure~\ref{fig:4steady} shows steady-state distributions obtained by
numerically solving Eq.~(\ref{FPsource}) with a source term centered
at $r_S=10\au$, steadily producing earth-mass planets at a rate
$\Lambda=(10^5\yr)^{-1}$. The topmost curve in each panel is the
steady-state distribution.  The enclosed curves, on the other hand,
are snapshots of the decaying distributions at selected time intervals
after the source is shut off.  We have verified that the steady-state
distributions for the MMSN in Panels I and III are fully consistent
with the analytic solution, Eqs.~(\ref{eq:mu})-(\ref{eq:fss}), for
power-law disks.  A fiducial normalization of $\epsilon=0.5$ for
torque fluctuations is assumed in all cases except in panel III, where
$\epsilon =0.02$ has been scaled down to extend the
migration-dominated regime out to $\sim 10\au$ in the MMSN.  Thus, at
$10\au$, the disks in Panels I and IV are diffusion-dominated and the
disks in Panels II and III are migration-dominated.

The distinction between diffusion-dominated and
migration-dominated evolutions is most clearly illustrated by
comparing panels~I and~III. While the initial steady-state peaked
distribution is eroded and advected for the most part in panel III,
with little diffusion at large radii, there is significantly more
diffusion at large radii in panel I. That distinction is not as clear
in the alpha disk models, however. Note that, in both panels~II
and~IV, the initial steady-state distribution is strongly peaked not
at the source radius, but at radii where migration
stalls (as seen from panel III in Fig.~\ref{fig:4dprop}). In addition,
even though diffusion is more important in the model with
$\alpha=10^{-3}$ in panel IV, the effect appears to be minor in
smoothing out only slightly the same sharp distribution features as in
panel II. This is partly explained by the strong contribution of
radially inhomogeneous diffusion to inward advection of planets, which
is always important in our alpha disk models (as explained in the
previous subsection). Even though diffusion is equally important at
$r_S=10\au$ in the models of panel I and IV (see $t_{\rm mig}/t_{\rm
diff}$ in Fig.~\ref{fig:4dprop}), it is more inhomogeneous in the alpha
disk model and thus promotes a significantly more efficient inward
advection of planets.

Our models with a burst of planet formation confirm these general
trends. To study this formation scenario, we solve the diffusion
equation (\ref{diffusion}) without source term and impose a specific
initial condition for the distribution of planets at $t=0$
instead. The initial distribution, $f(t\!=\!0,J)\!=\!f_i(J)$, is
modeled as a narrow Gaussian centered on $J_S=M_p\sqrt{GM_*r_S}$.
Figure~\ref{fig:4spike} shows the evolution of such distributions of
Earth-mass planets initially located at $r_S=10\au$ in the same four
disk models as in Fig.~\ref{fig:4steady}.

As before, a comparison between the two MMSN models shows that the
effects of diffusion are much more important in the model with
$\epsilon =0.5$ (panel I) than in the one with $\epsilon =0.02$ (panel
III). In particular, a sub-population of planets diffusing at large
radii is clearly visible in panel I. On the contrary, while diffusion
still acts to widen the initial distribution in panel III, the
dominant effect is a global advection of the distribution of planets.
Finally, in agreement with our previous results,
Figure~\ref{fig:4spike} shows that the most significant feature of
alpha disk models is the effect of rapid variations in the migration
rates of planets, which causes them to stall at specific locations in
the disk and results in time-varying peaks in the distributions of
panels II and IV. Even though diffusion is comparatively as important
in panels I and IV, at $10\au$, its effects appear to be much less
important in the alpha disk model of panel IV. We attribute this
difference, once again, to the more inhomogeneous diffusion in alpha
disk models, which contributes substantially to the inward advection
of planets into regions of the disk where diffusion becomes gradually
less important and the effects of stalling radii are felt more
strongly.

\subsection{Radial Excursions and Lifetimes}

\begin{figure}
\plotone{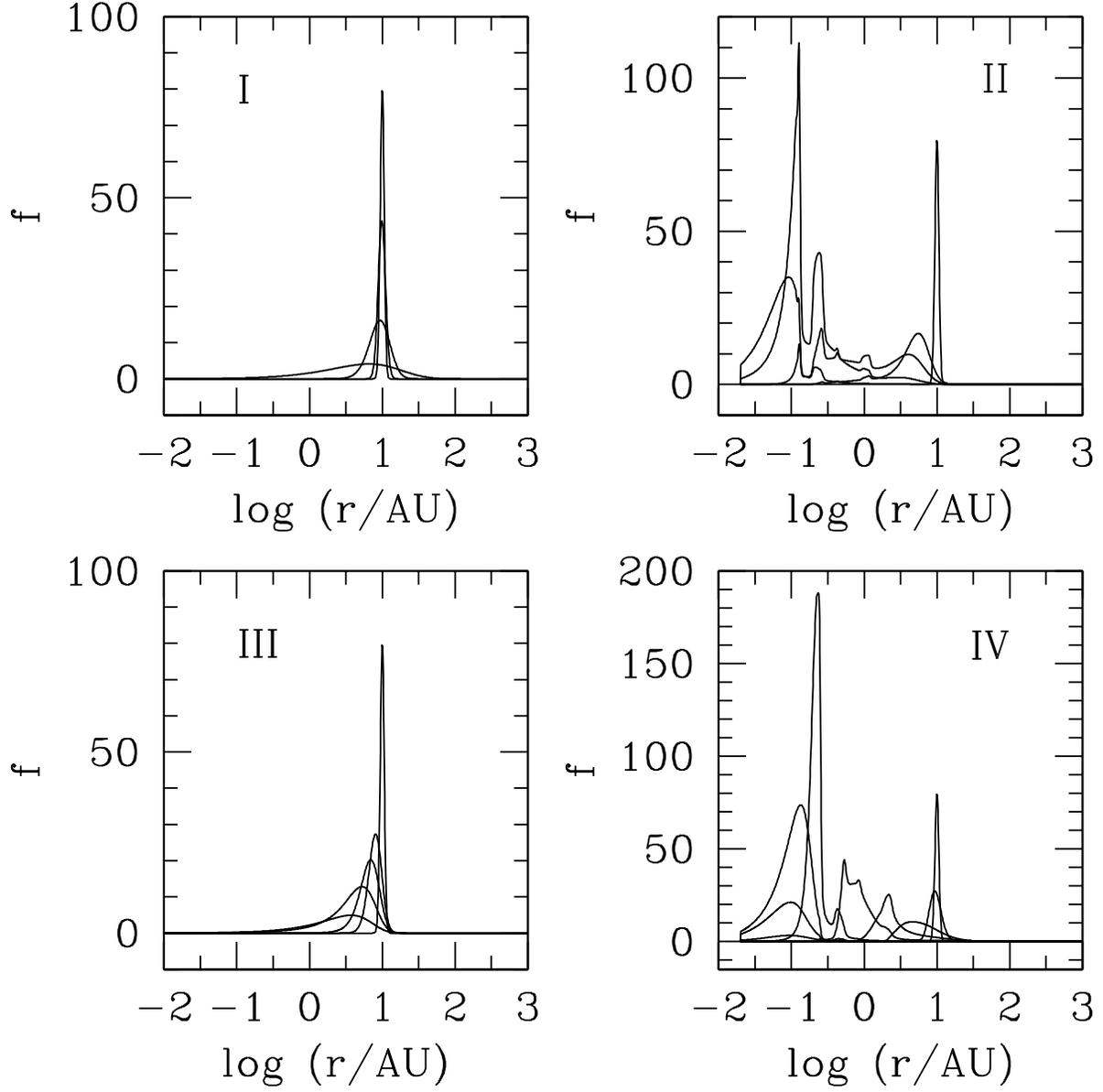}
\caption{Snapshots of the evolution of initially spiked distributions
corresponding to a burst of Earth-mass planet formation at
$r_S=10\au$.  The disk model in Panels I and III is the MMSN with
torque fluctuations normalized to $\epsilon=0.5$ and $0.02$,
respectively.  The disk models in Panels II and IV are alpha disks
with viscosity parameters $\alpha=0.02$ and $10^{-3}$, respectively
($\epsilon=0.5$).  Snapshots are shown at the following times after
burst formation, in log-years: (I) 4,5,6; (II) 6.8,7,7.2,7.4; (III)
5.8,6,6.2,6.4; (IV) 4,5,5.4,5.8,6.2,6.4,6.5,6.6.\label{fig:4spike}}
\end{figure}

Two important aspects of the orbital evolution of planets which are
not well captured by Figure~\ref{fig:4spike} are the properties of the
surviving minority of planets at late times and large radii,
and the representative planetary lifetimes in burst-formation models.

The fraction of planets that diffuse to orbital radii greater than
their formation radius, $r_S$, can be characterized by the
time-dependent quantity
\begin{equation}\label{eq:chia}
\chi_a(t)=\left.\int\limits_{J_a}^\infty f(t,J)dJ\right/
\int\limits_0^\infty f(0,J)dJ\,,
\end{equation}
where $J_a$ is the angular momentum at a fiducial radius $a[\au]>r_S$.
Hereafter, the integral over the initial distribution appearing in the
denominator of Eq.~(\ref{eq:chia}) will be normalized to unity.  The
maximum values of $\chi_{20}(t)$ and $\chi_{100}(t)$, reached
typically just after the majority of planets were accreted by the
star, are listed in Table \ref{tb:chia} for the four models of
Figure~\ref{fig:4spike}. Not surprisingly, models with a strong role
for diffusion tend to have a fairly large maximum fraction of planets
diffusing beyond $20\au$ ($\sim 10\%$) and $100\au$ ($\sim 1\%$),
while models with little diffusion have much reduced maximum fractions
of planets at large radii ($\lesssim 0.01\%$ beyond $20\au$, $\lesssim
10^{-6}$ beyond $100\au$).

\begin{table*}[ht]
\caption{Maximum fraction of planets beyond $a=20\au$ and $100\au$ in
the four models of Figure~\ref{fig:4spike}\label{tb:chia}}
\begin{tabular}{lccc}
Disk Model & Torque Fluctuations &$\chi_{20}^{\rm max}$ & $\chi_{100}^{\rm max}$ \\
\tableline
MMSN & $\epsilon=0.5$ & $1.4\times10^{-1}$ & $3.4\times10^{-2}$ \\
Alpha-Disk ($\alpha=0.02$) & $\epsilon=0.5$ & $1.9\times10^{-4}$ & $1.4\times10^{-6}$ \\
MMSN & $\epsilon=0.02$ & $1.0\times10^{-4}$ & $7.6\times10^{-8}$ \\
Alpha-Disk ($\alpha=10^{-3}$) & $\epsilon=0.5$  & $7.4\times10^{-2}$ & $9.7\times10^{-3}$ \\
\tableline
\end{tabular}
\end{table*}

We compute three additional time-dependent quantities to help us better
characterize the properties of surviving planets in our models. The first is
the survival probability, computed as
\[
P(t)= \int\limits_0^\infty\!f(t,J)dJ\,.
\]
The second is the median orbital radius of surviving planets, 
$r_{\rm med}(t)$, defined by
\[
\int\limits_0^{M_p\sqrt{GM_*r_{\rm med}(t)}} f(t,J)dJ= \frac{1}{2}P(t)\,.
\]
Finally, since the mean lifetime is often ill-defined (\S\ref{subsec:timedep}),
we calculate the median lifetime $t_m$ defined by $P(t_m)=1/2$.

\begin{figure}
\plotone{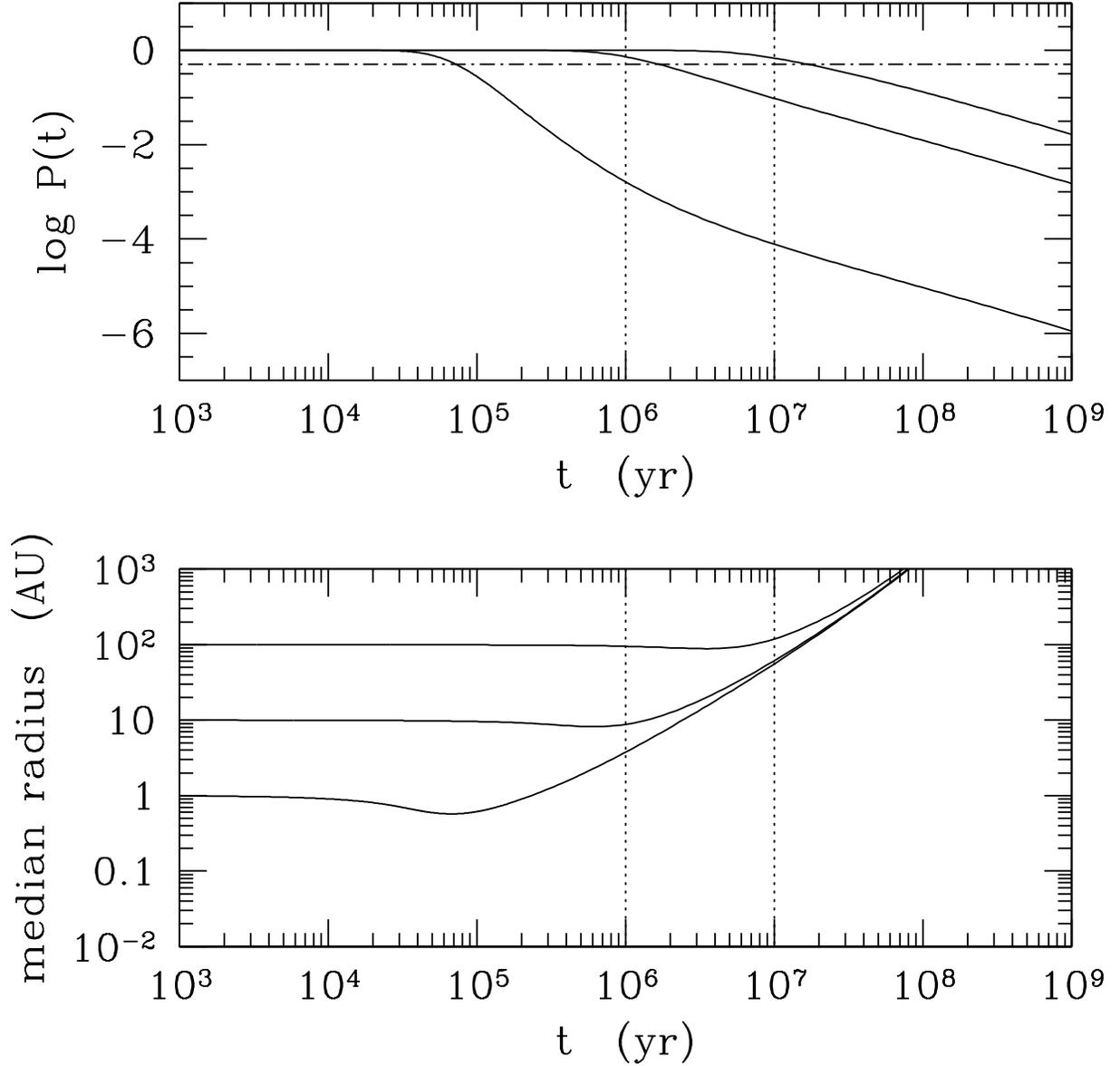}
\caption{Evolution of the survival probability, $P$ (upper panel), and
of the median radius of surviving planets (lower panel) for Earth-mass
planets formed in a localized burst at $t=0$ in the MMSN. Torque
fluctuations were normalized to $\epsilon =0.5$.  In each panel, the
three solid lines correspond to planets formed at $r_S=100$, $10$ and
$1\au$ (from top to bottom). Vertical dotted lines bracket typical
protoplanetary disk lifetimes. The horizontal dash-dotted line in the
upper panel corresponds to $P(t)=0.5$, which defines the median
lifetime of planets. A declining fraction of planets survives for long
times at increasingly large radii, in agreement with the self-similar
$P(t) \propto t ^{-1}$ scaling predicted
analytically.\label{fig:MMSNsv}}
\end{figure}

Figure~\ref{fig:MMSNsv} shows how the survival probability and the
median radius of surviving planets evolve with time in burst formation
models for Earth-mass planets in the MMSN. Three separate models with
burst formation at $r_S = 1$, $10$ and $100\au$ are shown for
comparison (solid lines). Note that the middle solid line corresponds
to the exact same model as the one shown in panel I of
Figure~\ref{fig:4spike}. For reference, vertical dotted lines bracket
typical lifetimes for gaseous proto-planetary disks
\citep[$10^6$--$10^7$~yrs; e.g.][]{hll01}. The median lifetime of the
population of planets, which is such that $P(t)=0.5$, is indicated by
a horizontal dash-dotted line in the upper panel. The lifetimes differ
substantially in the three models shown, from much shorter to somewhat
longer than the typical disk lifetimes. The self-similar $P(t) \propto
t^{-1}$ behavior predicted in \S\ref{subsec:timedep} is indeed
achieved at late times in the three models. As shown in the lower
panel of Figure~\ref{fig:MMSNsv}, the median radius of surviving
planets does not evolve much at early times and remains close to the
formation radius, as the distribution of planets mostly diffuses out
to both small and large radii. Once a significant fraction of planets
have been accreted by the central star, however, the surviving planets
are found at a median radius increasing almost quadratically with time
independently of where they were formed.  This is just what one would
expect from the self-similar solution (\ref{eq:ssim}): $r_{\rm
  med}^{1/2}\propto J_{\rm med}\to D_0 t\ln 2$.

\begin{figure}
\plotone{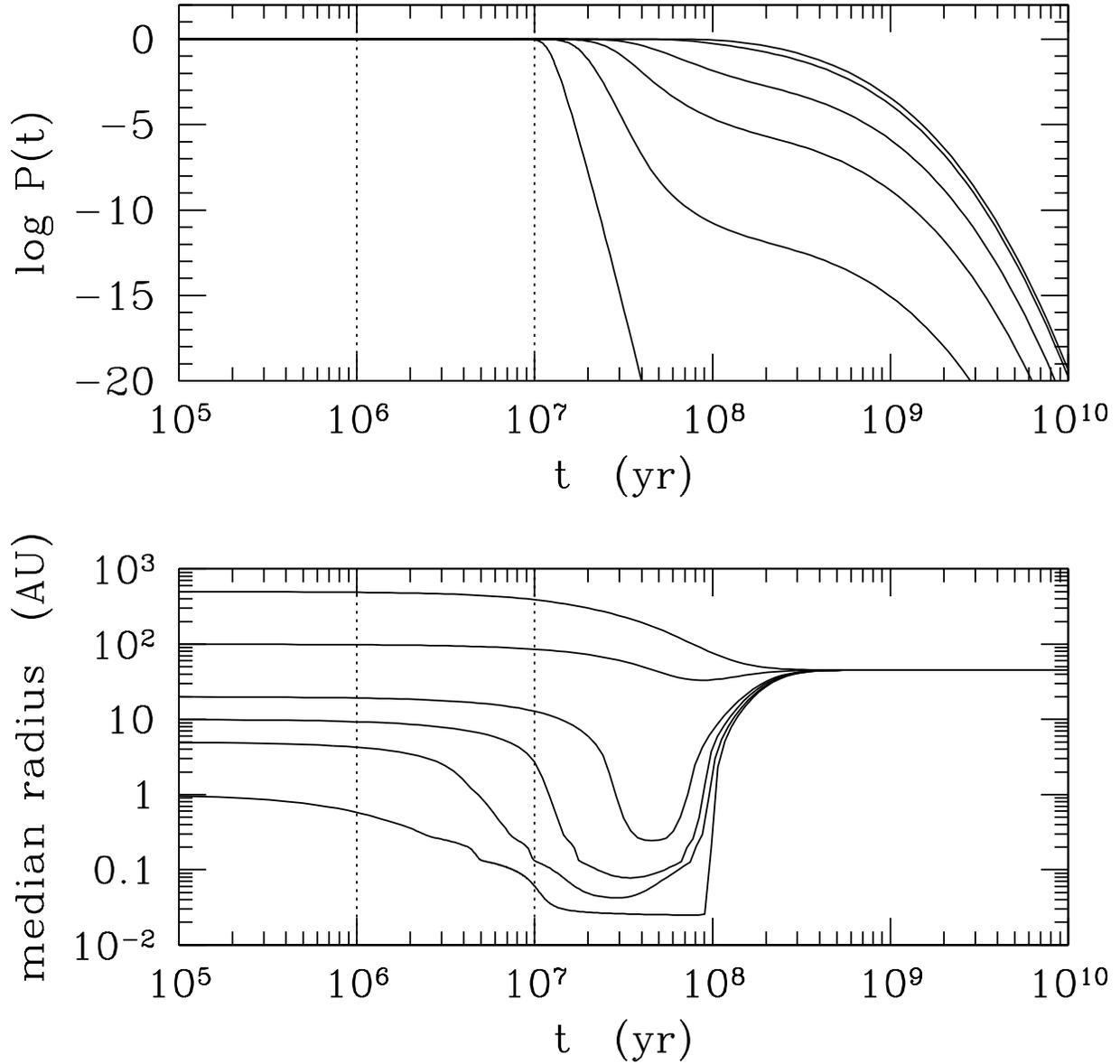}
\caption{Evolution of the survival probability, $P$ (upper panel), and
of the median radius of surviving planets (lower panel) for Earth-mass
planets formed in a localized burst at $t=0$ in the T-Tauri alpha-disk
model (with $\alpha=0.02$). Torque fluctuations were normalized to
$\epsilon =0.5$.  In each panel, the six solid lines correspond to
planets formed at $r_S=500$, $100$, $20$, $10$, $5$ and $1\au$ (from
top to bottom). Vertical dotted lines bracket typical protoplanetary
disk lifetimes.  As before, a declining fraction of planets survives
for long times but in these models the median radii reach an
asymptotic value at which the sum of the secular and diffusive
contributions to advection is minimized. The advection of most planets
at intermediate times is clearly visible in the median radius
evolution.
\label{fig:alpsv}}
\end{figure}

Figure~\ref{fig:alpsv} shows the evolution of the same two quantities
in burst formation models for Earth-mass planets in the alpha disk
(with $\alpha=0.02$) this time. Six separate models with burst
formation at $r_S = 1$, $5$, $10$, $20$, $100$ and $500\au$ are shown
for comparison (solid lines). Note that the fourth solid line from the
top corresponds to the exact same model as the one shown in panel II
of Figure~\ref{fig:4spike}. The results are qualitatively consistent
with the MMSN ones, with a few important differences.  There is no
self-similar behavior at late times in this case and the more dominant
role of advection in the alpha disk model tends to make the median
radius of surviving planets move inward initially, until the majority
of them are lost to the central star. Among the surviving minority,
however, $r_{\rm med}$ asymptotes to $\sim 50\au$ at late times,
independently of the formation radius, $r_S$.  This behavior is
explicable if, as predicted by the theorem cited in
\S\ref{subsubsec:eig}, the eigenmodes of Eq.~(\ref{diffusion}) are
discrete: then $f(t,J)$ converges upon the most slowly decaying mode,
and $r_{\rm med}(t)$ tends to the median radius of this eigenmode.
Figures~\ref{fig:4dprop} and~\ref{fig:alpsv} suggest that this radius
is close to that at which $t_{\rm mig}^{-1}+t_{\rm diff}^{-1}$ [more
accurately, $t_{\rm mig}^{-1}+t_{\rm adv}^{-1}$, where $t_{\rm
  adv}\equiv J/(\partial D/\partial J)$] is minimized, as might be
expected since planets should linger longest where the combined
effects of diffusion and migration are weakest. In the MMSN, both
timescales increase without limit as $J\to\infty$, and the spectrum of
eigenmodes is apparently continuous.

It is also worth emphasizing
that the largest peak value of $t_{\rm mig}$ at sub-AU distances in
alpha disk models (see panel III in Fig.~\ref{fig:4dprop}) effectively
limits the migration time of Earth-mass planets into their star to
values $\gtrsim 10^7$~yrs (for $\alpha=0.02$), which exceeds typical
disk lifetimes.

\begin{figure}
\plotone{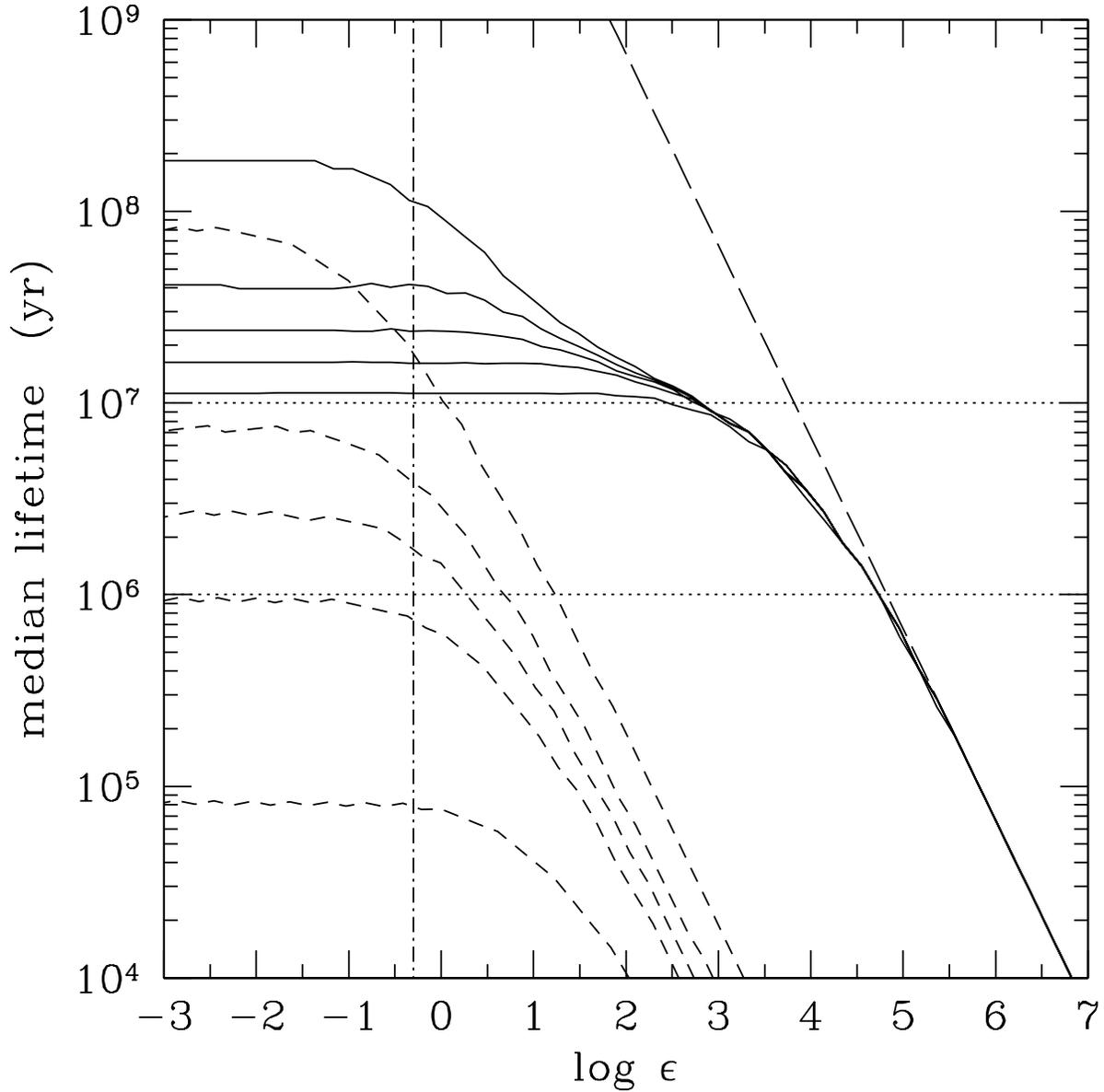}
\caption{Median lifetimes as a function of torque fluctuation
normalization, $\epsilon$, for Earth-mass planets formed in a burst at
radii $r_S=100$, $20$, $10$, $5$ and $1\au$ (from top to bottom) in
the MMSN (dashed lines) and in the alpha-disk model (solid lines;
$\alpha=0.02$). The vertical dash-dotted line corresponds to our
fiducial calibration of torque fluctuations ($\epsilon =
0.5$). Horizontal dotted lines bracket typical protoplanetary disk
lifetimes. The long-dashed line indicates the median lifetimes
expected in the alpha-disk model in the limit of negligible
mean torque.  Diffusion is unimportant at low $\epsilon$ values, but
it can significantly reduce the median lifetimes of planets formed at
any radius for large enough $\epsilon$ values.
\label{fig:medlrs}}
\end{figure}

Based on this general understanding of diffusive migration in our
burst-formation models, let us now focus more systematically on
several important model dependencies.  Figure~\ref{fig:medlrs}
illustrates how the median lifetimes of Earth-mass planets formed at
different radii in the MMSN and the alpha disk are affected by changes
in the normalization of torque fluctuations, $\epsilon$. For the
nominal $\epsilon =0.5$ value adopted (indicated by a vertical
dash-dotted line in Fig.~\ref{fig:medlrs}), diffusion mostly affects
the orbital evolution of planets formed at the largest radii, by
reducing somewhat their median lifetime. As $\epsilon$ is increased
above this nominal value, the trend becomes clearer. Median lifetimes
for all formation radii are considerably reduced at larger values of
$\epsilon$, more so in the MMSN than in the alpha disk.  The
long-dashed line in Figure~\ref{fig:medlrs} results when we neglect
the mean torque $\bar\Gamma$ in the governing equation
(\ref{diffusion}).  The convergence of the solid curves for the
alpha-disk model to a value independent of $r_S$ is related to
the value of the diffusion time, $t_{\rm diff}$, at the disk inner
boundary. For sufficiently large $\epsilon$, $t_{\rm diff}\ll
t_{\rm mig}$ everywhere on the grid.  But $t_{\rm diff}$ increases
toward small radius in the alpha disks (Fig.~\ref{fig:4dprop}).  So
the median lifetime in the large-$\epsilon$ regime is comparable to
$t_{\rm diff}$ at the inner boundary, $r_{\rm min}$; we have verified
that $t_{\rm med}$ decreases as $r_{\rm min}$ is increased.
Nevertheless, in all cases, the effect of diffusion is to reduce the
median lifetime.

\begin{figure}
\plotone{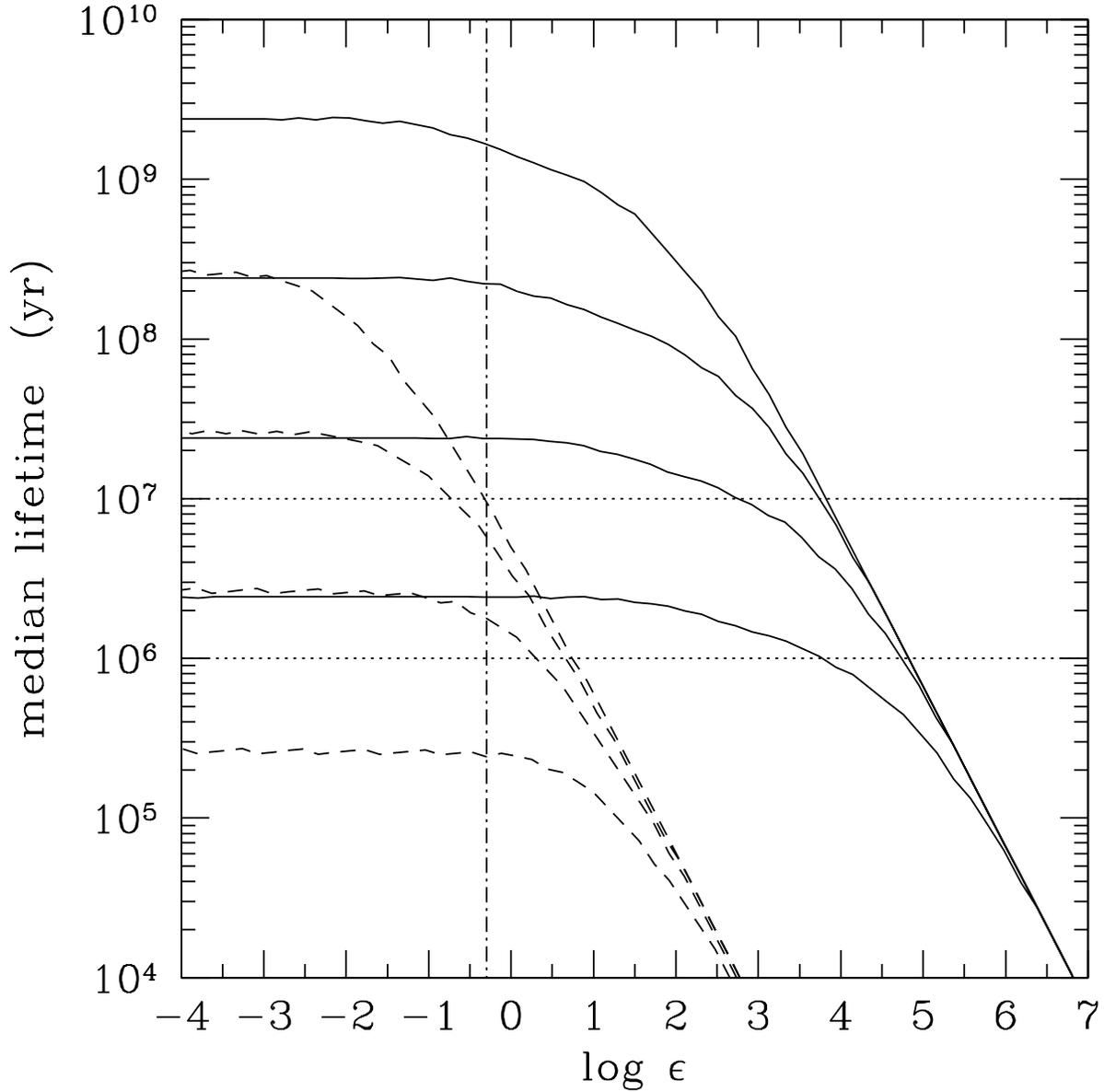}
\caption{Median lifetimes as a function of torque fluctuation
normalization, $\epsilon$, for planets of various masses all formed in
a burst at $r_S=10\au$ in the MMSN (dashed lines) and in the
alpha-disk model (solid lines; $\alpha=0.02$). The various curves
correspond to planets of mass $M_p=0.01$, $0.1$, $1$ and $10 M_\oplus$
(from top to bottom).  The vertical dash-dotted line corresponds to
our fiducial calibration of torque fluctuations ($\epsilon =
0.5$). Horizontal dotted lines bracket typical protoplanetary disk
lifetimes. Diffusion is unimportant at low $\epsilon$ values, but it
can significantly reduce the median lifetimes of planets of any mass
($\lesssim 10 M_\oplus$) for large enough $\epsilon$ values.
\label{fig:medlMp}}
\end{figure}

Figure~\ref{fig:medlMp} illustrates how the median lifetimes of
planets of various masses, all formed at $r_S=10\au$ in the MMSN and
in the alpha disk, are affected by changes in the normalization of
torque fluctuations, $\epsilon$. For the nominal $\epsilon =0.5$
value, diffusion mostly affects the orbital evolution of the lower
mass planets, and reduces somewhat their median lifetimes. This
results from the larger values of $t_{\rm mig}$ at lower planet mass,
whereas $t_{\rm diff}$ is independent of planet mass.  As
$\epsilon$ is increased, diffusion progressively dominates and the
distinction among planets of various masses gradually disappears.
For large enough values of $\epsilon$, the median lifetime of all
planets with masses $M_p \lesssim 10 M_\oplus$ formed at $r_S=10 \au$ is
reduced below typical disk lifetimes. As in the previous figure, however,
the lifetimes for $\epsilon\gg 1$ in the alpha disk
are dominated by the diffusion time at the inner boundary.

\section{Discussion and Conclusion} 

As anticipated by \citet{np04}, \citet{lsa04}, and \citet{nelson05},
our Fokker-Planck treatment confirms that diffusion driven by
turbulent torque fluctuations in proto-planetary disks can greatly
influence the orbital evolution of embedded low-mass proto-planets. Contrary
to some expectations, diffusion does not promote planetary survival but
instead systematically reduces the lifetime of most planets in the
disk. However, it does help a small fraction of planets diffuse to
large radii where they can survive for extended periods. (We consider
that a planet ``survives'' if its orbit remains larger
than the inner edge of the disk.)

Our results point to a potentially important role for
non-deterministic effects in planet formation scenarios. In models
with a significant level of diffusion, most proto-planets in the
Earth-mass range could end up being accreted by their stars, while
only a small fraction of them would survive by diffusing to large
radii. In scenarios where all proto-planetary disks are efficient at
forming planets, this could be interpreted as leading to only a
fraction of all potential host stars being successful in keeping a
system of surviving planets. Even systems starting with very similar
initial conditions may end up with very different planetary orbital
configurations once their gaseous disks disappear.

Neglecting the effects of direct gravitational interactions between
proto-planets, which are likely to happen from differential rates of
migration and diffusion, is one of many model limitations in our
work. We have already mentioned that our study is restricted to
proto-planets of low enough masses that they do not affect in any way
the structure of their host disk. This excludes the late stages of
giant planet formation, which are obviously of considerable
interest. In particular, it has been suggested that migration and
diffusion, each independently, could accelerate the rate of growth of
cores until they reach the critical mass at which runaway envelope
accretion occurs \citep[e.g.][]{ra03,al05a,al05b}. Studying these
possibilities within the context of global models such as ours would
be interesting.

Even within the strict regime of applicability of our models,
significant uncertainties exist. Let us mention a few. We have focused
our work on strictly circular orbits. We have assumed that the
underlying proto-planetary disk structure does not evolve with time,
and that the amplitude $\delta\Sigma/\Sigma$ of surface-density
fluctuations is constant with radius. We have assumed that torque
fluctuations can simply be added to a laminar mean torque which is not
modified by the turbulence. The laminar torque itself is calculated
from a simple extension of the two-dimensional Lindblad torque theory,
which does not account for the role of corotation torques.

Despite all these uncertainties, we believe that our results should be
broadly valid, provided that turbulence and torque fluctuations in
proto-planetary disks satisfy the various stochasticity criteria
enunciated in \S\ref{sec:FP}. Understanding better the nature of
turbulence in proto-planetary disks and in particular indications for
long-term correlations in MHD simulations may turn out to be crucial
in this context. We would like to issue a plea to the simulation
community to study these questions, and in particular, whether the
dimensionless amplitude and correlation time of density and torque
fluctuations depend only upon the strength of the turbulence
(\emph{via} $\alpha$) or upon the thickness $h/r$ as well
(\S\ref{subsec:calib}).  The answers will matter both quantitatively
(in determining the relative importance of migration \emph{vs.}
diffusion in global models such as ours) and qualitatively (in
justifying a Fokker-Planck approach or in falsifying it).

We have seen that details of the protoplanetary disk structure (MMSN
\emph{vs.} alpha disks) do influence our results at the quantitative
level. In that respect, the possibility that a magnetically layered
structure with quiescent dead zones exists in weakly-ionized regions
of proto-planetary disks \citep{g96,gni97,bh98,ftb02,in06} may be
important. Indeed, the results of \citet{fs03} suggest that the
nature of turbulence in dead zones differs qualitatively from that in
MHD turbulent regions, perhaps leading to comparatively reduced torque
fluctuations \citep[see also][for other suggested effects of dead
zones in planet formation scenarios]{mp03}.

The general tendency for a fraction of planets to diffuse to large
radii in proto-planetary disks may have consequences of some
observational relevance. We note, for instance, that \citet{va04} have
looked into scenarios for outward planet migration in an attempt to
explain observed structures in dust disks as resulting from
perturbations by planets located at radii beyond those at which they
are thought to be able to form. Rather than invoking outward
migration, this may be expected in scenarios with a prominent role for
diffusion, as illustrated by our various models.

\section*{Acknowledgments} 

We thank Richard Nelson for help with understanding his simulations.
This material is based on work supported in part by the National
Aeronautics and Space Administration under Grant NAG5-11664 issued
through the Office of Space Science.

\clearpage

\end{document}